\documentclass{PNASTWO}

\usepackage{graphicx}

\usepackage{amssymb,amsfonts,amsmath}
\usepackage{caption}
\usepackage{prettyref}
\newrefformat{sfig}{SI Fig. \ref{#1}}
\newrefformat{tab}{SI Table \ref{#1}}
\DeclareCaptionLabelFormat{support}{SI #1 #2}
\pdfoutput=1
\begin{document}

%%%%%%%%%%%%%%%%%%%%%%%%%%%%%%

%% You may leave the \conflictofinterest and \track commands with the %% default values as PNAS will collect this information during the %% online submission process.

\footcomment{Author contributions:
RWC and FG designed research; JK and AFMM developed simulation code; JK and TH performed research; JK, TH, RWC and FG analyzed data; JK, AFMM and FG wrote the paper.}

\conflictofinterest{Conflict of interest footnote placeholder} 

\track{Insert 'This paper was submitted directly to the PNAS office.' when applicable.}

\footcomment{Abbreviations: $N$, N-cadherin; $E$, E-cadherin;  $C$ cell, cone cell; $c$ cell, mutant cone cell;  $P$ cell, primary pigment cell; $p$ cell, mutant primary pigment cell} 

%% If you wish to include abbreviations (or any other footnote that doesn't %% use a footnote marker), you may define these via the \footcomment command. 

%\footcomment{Abbreviations: SAM, self-assembled monolayer; % OTS, octadecyltrichlorosilane}

%% For titles, only capitalize the first letter %% \title{Almost sharp fronts for the surface quasi-geostrophic equation} 

\title{Cell adhesion and cortex contractility determine cell patterning in the Drosophila retina}

%% Enter authors via the \author command. Use \thanks to indicate the %% corresponding author. Use \affil to define affiliations. 

\author{Jos K\"afer\thanks{To whom correspondence should be addressed. E-mail: jkafer@spectro.ujf-grenoble.fr}\affil{1}{Laboratoire de Spectrom\'etrie Physique, UMR 5588, UJF Grenoble I \& CNRS, 140 Avenue de la Physique, 38402 Saint Martin d'H\`eres, France}, Takashi Hayashi\thanks{Current address: Department of Zoology, Michigan State University, 203 Natural Science, East Lansing, MI, 48824, USA}\affil{2}{Department of Biochemistry, Molecular Biology and Cell Biology, Northwestern University, Evanston, Illinois 60208, USA}\affil{3}{Department of Biophysics and Biochemistry, Graduate School of Science, University of Tokyo, Tokyo 113-0033, Japan}, Athanasius F.M. Mar\'ee\affil{4}{Theoretical Biology/Bioinformatics, Utrecht University, Padualaan 8, 3584 CH Utrecht, the Netherlands}, Richard W. Carthew\affil{2}{}, \and
Fran\c{c}ois Graner\affil{1}{}}

%% You may leave the \contributor command with the default value. 
\contributor{Submitted to Proceedings of the National Academy of Sciences of the United States of America}

%% The \maketitle command is necessary to build the title page.

 \maketitle

%%%%%%%%%%%%%%%%%%%%%%%%%%%%%%%%%%%%%%%%%%%%%%%%%%%%%%%%%%%%%%%% 
\begin{article}

\begin{abstract}
Hayashi and Carthew (Nature 431 {[}2004], 647) have shown that the packing of cone cells in the \emph{Drosophila} retina resembles soap bubble packing, and that changing E- and N-cadherin expression can change this packing, as well as cell shape.

The analogy with bubbles suggests that cell packing is driven by surface minimization. We find that this assumption is insufficient to model the experimentally observed shapes and packing of the cells based on their cadherin expression. We then consider a model in which adhesion leads to a surface increase, balanced by cell cortex contraction. Using the experimentally observed distributions of E- and N-cadherin, we simulate the packing and cell shapes in the wildtype eye. Furthermore, by changing only the corresponding parameters, this model can describe the mutants with different numbers of cells, or changes in cadherin expression. 
\end{abstract}

%% When adding keywords, separate each term with a straight line: | 

%% The first letter of the article should be drop cap: \dropcap{} 

\dropcap{C}ell adhesion molecules are necessary to form a coherent multicellular organism. They not only hold cells together, but differential expression of different types of these molecules plays a central role during development. Members of the cadherin family are the most widespread molecules that mediate adhesion between animal cells, and their role has been demonstrated in cell sorting, migration, tumor invasibility, cell intercalation, packing of epithelial cells, axon outgrowth and many more \cite{Steinberg.s63,Niewiadomska.jcb99,Foty.ijdb04,Lecuit.tcb05,Classen.dc05,Gumbiner.nrmc05}. 
We here focus on the role of adhesion in the determination of epithelial cell shape \cite{Carthew.cogd05}.

In the compound eye of \emph{Drosophila}, the basic unit, the ommatidium, is repeated approximately $800$ times. All ommatidia have the same cell packing, which is essential for correct vision. The ommatidium consists of four cone cells, which are surrounded by two larger primary pigment cells. These `units' are embedded in a hexagonal matrix, constituted by secondary and tertiary pigment cells, and bristles (c.f. \prettyref{sfig:confocal}, \cite{Wolff.ddm93}).

Two of us \cite{Hayashi.n04} showed that cadherin expression influences ommatidial cone cell packing. 
Two cadherin types, E- and N-cadherin, are expressed in different cells: all interfaces bear E-cadherin, while N-cadherin is present only at interfaces between the four cone cells (\prettyref{sfig:confocal}). Cadherin-containing adherens junctions form a zone close to the apical cell surface, allowing the retina epithelium to be treated as a 2D tissue.
In the wildtype and in {\em Roi}-mutant ommatidia with two to six cone cells, these cone cells assume a packing (or topology, that is, relative positions of cells)  strikingly similar to that of a soap bubble cluster.
When cadherin expression is changed in a few or all of the cells, the topology can change. More frequently, only the geometry (individual cell shapes, contact angles at the vertices, interface lengths) changes.

The soap films between bubbles are always under a positive tension, $\gamma>0$. This surface tension describes the energy cost of a unit of interface between bubbles, and drives their packing.
At equilibrium, in a 2D foam layer, soap bubbles meet by three at each vertex, since four-bubble vertices are unstable \cite{Plateau.setl73,Weaire.pf99}. In addition, since $\gamma$ is constant and the same for all interfaces, bubble walls meet at equal ({\it i.e.} 120$^\circ$) angles. More precisely, the surface energy (or rather the perimeter energy, for a 2D foam) is $\gamma P$, where $P$ is the total perimeter of soap films. The foam reaches equilibrium when it minimises $P$ (since $\gamma$ is constant), balanced by another constraint fixing each bubble's area.

It has been proposed that cells minimize their surface, like soap bubbles \cite{Thompson.ogaf42,Chichilnisky.jtb86,Lecuit.nrmc07}. Since the surface mechanics of bubbles are quite simple, they can easily be described in a model. However, calculating the equilibrium shape of a cluster of more than four bubbles is difficult \cite{Chichilnisky.jtb86}; for this purpose, we use a numerical method \cite{Graner.prl92,Glazier.pres93}, in order to test if cell patterning is based on surface minimization. Here, the only biological ingredient is differential adhesion \cite{Steinberg.s63,Chichilnisky.jtb86,Graner.jtb93a}: an interface between two cells has a constant tension, that is lower when the adhesion is stronger \cite{Graner.prl92,Glazier.pres93}.

Cells, however, differ greatly from bubbles, both in their membrane and internal composition. Surface tension has been shown to be determined up to a large extent by the cortical cytoskeleton \cite{Sheetz.tcb96,Raucher.bj99,Dai.bj99,Morris.jmb01}. Adhesive cells have a tendency to increase their contact interfaces \cite{Thoumine.ebj99}, not to minimize them. Lecuit and Lenne \cite{Lecuit.nrmc07} recently reviewed a large number of experiments, and show that a cell's surface tension results from the opposite actions of adhesion and cytoskeletal contraction. These are the ingredients of a second model \cite{Graner.jtb93b,Ouchi.pa03}.

Our approach is to find out if the observed cell packings and shapes can be described with one of these models, based on the knowledge we have from the experiments. With minimal and realistic assumptions, only the second model reproduces the topology and geometry of the wildtype and mutant ommatidia.

This shows that the competition between adhesion and cell cortex tension is needed to describe this specific cell pattern. We thus confirm and refine the conclusion that surface mechanics are involved in the establishment of cell topology and geometry. Adhesion plays an important role therein, but its role can only be understood when taking into account its effect on the cortical cytoskeleton. 

\section{Results}

\subsection{Model simulations}

Why certain shapes are observed more often than others depends on the developmental history of the tissue, which is determined by e.g. the sequence of cell differentiation, and cell divisions and deaths. Since a lot is still unknown about the developmental history, we do not include it in the modeling. However, since cells seem in mechanical equilibrium at any moment in development (c.f. \cite{Lecuit.nrmc07}), future insights in developmental gene regulation could be translated in parameter changes that permit the modeling of the dynamics of development.

Simulations thus start from unstable initial conditions (\prettyref{sfig:Initial-conditions}) designed to favor the random search 
of final, stable topologies. We do not expect to find a quantitative correspondence between the frequency of topologies in simulations and experiments.
We regard only the final result of the model simulations: we have found a local equilibrium, when the simulated shape does not change anymore. 

We compare this shape with the experimental results (topology, geometry). Distinguishing between topologies is trivial. But, due to the variability of membrane fluctuations,  we find that it is difficult to describe the geometrical characteristics (e.g. contact angles for the mutant ommatidia, interface lengths, elongation of cells) by quantitative measurements: one obtains more information by looking at the image (`eyeballing'). Quantitative measurements serve as a complement to the eyeballing when enough data are available (c.f. Fig.~\ref{fig:wildtype}), not as replacement. We determine for each model which parameters do influence the shape of the cone cells; for the other parameters, we choose reasonable values (e.g., a compromise between simulation speed and precision, c.f. \cite{Maree.scbm07}). 

We assume that (i) the adhesion strength is determined by the presence of these cadherins: when the two of them are present (i.e. at interfaces between cone cells), adhesion is thus stronger. Mutants should be modeled by only changing existing parameters. We thus require that (ii) to model the {\it Roi}-mutants, we only need to change the number of cone cells; (iii) to model the cadherin mutant ommatidia, only the adhesion for the mutant cells should be changed (i.e. diminished for deletion, increased for overexpression); and (iv) all cells of a cell type that share the same mutation should be modeled using the same parameter values.

\subsection{Constant tension model}

A stronger adhesion between cells $i$ and $j$ is represented by a lower interfacial tension \cite{Chichilnisky.jtb86,Graner.jtb93a,Graner.prl92,Glazier.pres93}, 
$\gamma_{ij} \geq0$, which is a constant depending only on the cell types of $i$ and $j$.
We minimise the energy:\begin{equation} {\cal E}=\sum_{\mbox{\scriptsize{interfaces}}}\gamma_{ij}P_{ij}+\lambda_{A}\sum_{\mbox{\scriptsize{cells}}}\left(A_{i}-A_{0i}\right)^{2}\,.\label{eq:standard hamiltonian}\end{equation} 
$P_{ij}$ is the length of the interface between cells $i$ and $j$, $A_{i}$ is the cell's area (the 2D equivalent of volume), $A_{0i}$ is the cell's preferred area (target area), and $\lambda_{A}$ is the area modulus (a lower value allows more deviations from $A_{0}$).
The values of $A_{0i}$ are inferred from the experimental pictures, with cone ($C$) cells being smaller than primary pigment ($P$) cells. 

We assume $C$-$C$ adhesion $\gamma_{CC}$, mediated by both E- and N-cadherin, to be stronger than $C$-$P$ and $P$-$P$ adhesion, $\gamma_{CP}$ and $\gamma_{PP}$, which are mediated by E-cadherin alone. We assume the latter two to be equal: $\gamma_{CC}<\gamma_{CP}=\gamma_{PP}$. 
Only three parameters need to be explored extensively: 
$\gamma_{CC}$, $\gamma_{CP}$ ($=\gamma_{PP}$), and $\lambda_{A}$. The tensions $\gamma$ influence the cell shapes directly, whereas $\lambda_{A}$ determines a cell's deviations from the target area. 

\begin{figure}
\includegraphics{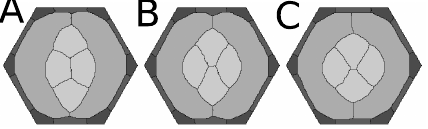}
\caption{Constant tension model
simulations. \textbf{A}: $\gamma_{CC}=40$, $\gamma_{CP}=\gamma_{PP}=80$. \textbf{B}: Same as (A), but with lower tension (stronger adhesion) between the polar and equatorial cone cell, $\gamma_{\mbox{\scriptsize{polar, equatorial}}}=20$. \textbf{C}: Same as (A), but with lower tension (stronger adhesion) between the primary pigment cells: $\gamma_{PP}=40$.}\label{fig:tension} \end{figure}

Starting the simulations with a four-cell vertex (\prettyref{sfig:Initial-conditions}A), we systematically find an incorrect topology (Fig.~\ref{fig:tension}A): the anterior and posterior $C$ cell touch. 
Even if we force the correct one, where the polar and equatorial $C$ cell touch, it is unstable and decays into the incorrect one: the interfaces between the $P$ cells are under tension, and pull the  
polar and equatorial $C$ cells apart.

To obtain the correct topology, we need another assumption: either that the adhesion between polar and equatorial $C$ cells is stronger
(Fig.~\ref{fig:tension}B); or that the $P$ cells pull less on them (by having a stronger adhesion, Fig.~\ref{fig:tension}C). 
Still, the geometry is quite different from the experiments: notably the interface between the polar and equatorial $C$ cell is too short in simulations. Besides, there is no experimental evidence to support these assumptions.

Another optimization strategy is to determine (up to a prefactor) the tensions of three interfaces A, B, D that meet in a vertex from the experimentally observed contact angles $\alpha$, $\beta$, $\delta$ ($\alpha+\beta+\delta=360^\circ$) by 
 using $\gamma_{A}/\sin\alpha=\gamma_{B}/\sin\beta=\gamma_{D}/\sin\delta$ \cite{Langmuir.jcp33,Thompson.ogaf42}.
We inject those tensions in the model. By construction we obtain the correct contact angles, and thus topology; but the overall geometry (especially the interface lengths) differs considerably from observations (results not shown).

For the mutant ommatidia, the requirements (ii) to (iv) mentioned above could not be satisfied with this model: there are too many cases where other parameters need to be changed as well. We conclude that this model is insufficient to coherently describe the experiments.

To obtain the observed shapes, it would certainly be possible to choose a tension for each individual interface. But if the tension was just an input parameter without biological basis, then the model would not be predictive, nor help to understand the differences between the cells. 
 
\subsection{Variable tension model}

Adhesion between two cells tends to extend their contact length; it thus contributes negatively to the energy, $-J_{ij}P_{ij}$, where $J>0$: in agreement with intuition, a higher $J$ describes a stronger adhesion, while $J=0$ in absence of adhesion \cite{Graner.jtb93b,Ouchi.pa03}. 

This extension is compensated by an elastic cell cortex term, $\lambda_{P}\left(P_{i}-P_{0i}\right)^{2}$, where $\lambda_{P}$ is the perimeter modulus, and $P_{0i}$ is the target perimeter of cell $i$. The cell perimeter is the sum of its interfaces, $P_{i}=\sum_{j}P_{ij}$. 
We thus minimise the energy:\begin{equation} {\cal E}=-\sum_{\mbox{\scriptsize{interfaces}}}J_{ij}P_{ij}+\sum_{\mbox{\scriptsize{cells}}}\left[\lambda_{P}\left(P_{i}-P_{0i}\right)^{2}+\lambda_{A}\left(A_{i}-A_{0i}\right)^{2}\right]\,.\label{eq:adhesion hamiltonian}\end{equation} 

The interfacial tension  $\gamma_{ij} = \partial {\cal E}_{ij}/\partial P_{ij}$
between cells $i$ and $j$ is the energy change associated with a change in membrane length (c.f. \cite{Lecuit.nrmc07}); eq.~\eqref{eq:adhesion hamiltonian} yields: \begin{equation}
\gamma_{ij}=-J_{ij}+2\lambda_{P}\left(P_{i}-P_{0i}\right)+2\lambda_{P}\left(P_{j}-P_{0j}\right)\,.\label{eq:tension}\end{equation} 
As in the previous model, $\gamma_{ij}$ is positive, else the cell would be unstable. However, it is  no longer an input parameter. A stronger adhesion (high $J$) decreases the tension: this will usually cause an extension of the perimeter, which increases this tension even more.

We represent all adhesion terms as combinations of E- and N-cadherin mediated adhesion ($J_{E}$ and $J_{N}$, respectively). In the wildtype, the adhesion between $C$ cells is mediated by both cadherins, so $J_{CC}=J_{E}+J_{N}$; whereas all other interfaces only have E-cadherin, so $J_{PP}=J_{CP}=J_{E}$.  
Values of $A_{0}$ are estimated from pictures.
The target perimeter $P_{0}$ (expressed in units of $2\sqrt{\pi A_{0}}$) should be larger for cells that deviate more from a circular shape, i.e. for the $P$ cells.   
  
We thus adjust 6 main parameters:
$J_{E}$, $J_{N}$, $P_{0C}$, $P_{0P}$, $\lambda_{P}$, $\lambda_{A}$, which is too much to explore systematically. We adjust the parameters by hand, for wildtype and mutant configurations simultaneously, since the wildtype alone does not sufficiently constrain the number of optimal parameter combinations.

Unless indicated, throughout this paper, and for all figures except Fig.~\ref{fig:tension}, we use eq. \eqref{eq:adhesion hamiltonian} with the same set of parameters (\prettyref{tab:Simulation-parameters}) for  wildtype and mutants. 

\subsection{Wildtype}

Starting the simulations with a four-cell vertex (\prettyref{sfig:Initial-conditions}A), the cells relax  either into the correct topology where the polar and equatorial cells touch 
(Fig.~\ref{fig:wildtype}); or into the incorrect one where anterior and posterior cells touch (analogous to Fig. \ref{fig:tension}A). Both topologies are stable, i.e. they are local energy minima.

In the correct topology, the geometry of the simulated ommatidium resembles well the experimental pictures. More quantitatively,  the contact angles measured in simulations and in experiments agree as well (Fig.~\ref{fig:wildtype}). In contrast to the constant tension model, we do not need additional assumptions.

We found that the adhesion of secondary and tertiary pigment cells should be much stronger than can be expected from E-cadherin alone ($J_{23}>J_{E}$, \prettyref{tab:Simulation-parameters}), otherwise they loose contact. Experimentally, deleting the E-cadherin of these cells does not induce any geometrical or topological change \cite{Hayashi.n04}. Both experiments and simulations thus suggest that secondary and tertiary pigment cells might have other adhesion molecules than E- and N-cadherin.

\subsection{{\it Roi}-mutants}

Without any additional parameter, we can simulate different numbers of $C$ cells (\emph{Roi}-mutants); the total size of the simulation lattice is adjusted accordingly. For one, two, three and five $C$ cells, only one topology is observed in experiments, and the same one in simulations (\prettyref{sfig:Roi}). 

\begin{figure}
\includegraphics{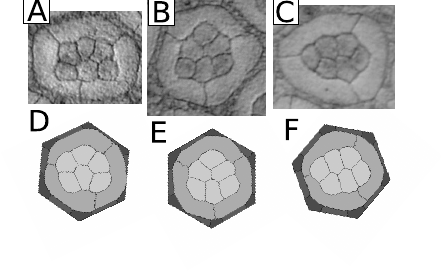}
\caption{Roi mutants with 6 cone cells.
\textbf{A-C}: Experimental pictures, from ref. \cite{Hayashi.n04}. \textbf{D-F}: Corresponding simulations.}\label{fig:Roi} \end{figure}

For six $C$ cells, three topologies are observed experimentally (Fig.~\ref{fig:Roi}A-C). 
Theoretically, there are two more possible equilibrium topologies for 6-cell aggregates, which are never observed although one of them has a smaller total interface length (simulations using the Surface Evolver, S. Cox, unpublished results 2004). We here performed a total of $42$ Potts model simulations with different random seeds (see Methods), and found only three topologies (Fig.~\ref{fig:Roi}D-F): they correspond to the observed ones.

We observe in  Fig.~\ref{fig:Roi}A and C that the entire ommatidium is elongated. Besides, ommatidia of \emph{Roi}-mutants do not all have six sides and are assembled into a disordered pattern (see \cite{Hayashi.n04}). Thus, in \emph{Roi}-mutants, ommatidia have variable shapes, which origin is not easily understood (especially for mutants with more pigment cells). Since in turn the shape of the ommatidium influences the geometry of its $C$ cells (results not shown), studying the geometry of the $C$ cells in more details would only by possible by adding more free parameters.

\subsection{N-cadherin mutants}

Again without any additional parameter, simply by suppressing  $J_{N}$, we could predict the pattern of ommatidia with N-cadherin deficient $C$ cells. Since 
N-cadherin is only present on interfaces between $C$ cells, deletion means we set the adhesion between mutant and wildtype $C$ cells as $J_{cC}=J_{cc}=J_{E}$ (mutant cells are denoted by lower case letters).

\begin{figure*}
\includegraphics{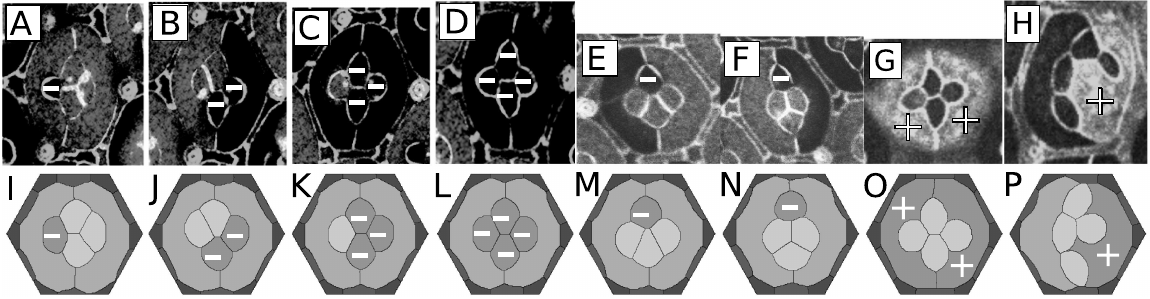}
\caption{N-cadherin mutants. Mutant
cells are indicated with a {}``+'' for overexpression, {}``-'' for deletion. \textbf{A-H}: Experimental pictures, A-D from ref. \cite{Hayashi.n04}. \textbf{I-P}: Corresponding simulations.}\label{fig:Ncad}
\end{figure*}

We predict the correct topologies (Fig.~\ref{fig:Ncad}A-F and I-N), most of which are the same as in the wildtype. We predict qualitatively the main geometrical differences between mutants and wildtype:
(i) the length of the interfaces between mutant cells and wildtype $C$ cells decreases; (ii) the contact angles change; (iii) the interface length between the remaining wildtype $C$ cells increases (Fig.~\ref{fig:Ncad}A-B and I-J); and (iv) the length of the central interface increases (Fig.~\ref{fig:Ncad}D and L).

When the polar or equatorial cell is the only $C$ cell without N-cadherin, 
we simulate (Fig.~\ref{fig:Ncad}M-N) both topologies that coexist in experiments (Fig.~\ref{fig:Ncad}E-F). 
 
To simulate one mutant $P$ cell that mis-expresses N-cadherin, we optimize $J_{Cp}$. While for the wildtype $J_{CP}=J_{E}=150$, we find an increase for the mutant, $J_{Cp}=150+600$.
The high adhesion of this $P$ cell with the $C$ cells severely disrupts the normal configuration. Many topologies that differ considerably form the wildtype are observed in experiments and simulations (e.g. Fig.~\ref{fig:Ncad}H,P).
When both $P$ cells mis-express N-cadherin, they balance each other and the topology is back to normal (Fig.~\ref{fig:Ncad}G). Optimization yields $J_{pp}=150+700$ (Fig.~\ref{fig:Ncad}O and  \prettyref{sfig:sens2}). Both $J_{Cp}$ and $J_{pp}$ are higher than the wildtype value of $C$-$C$ adhesion ($J_{CC}=J_{E}+J_{N}=150+450$).

\subsection{E-cadherin mutants}

\begin{figure}
\includegraphics{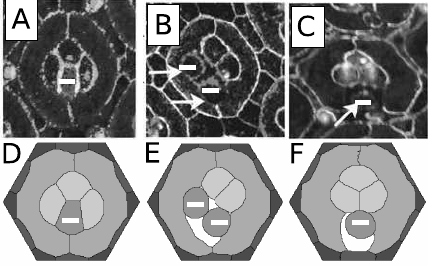}
\caption{Loss of adhesion. Mutant
cells are indicated with a {}``-''. \textbf{A}: A mutant cone cell lacking E-cadherin. \textbf{B-C}: Double mutant cone cells for E-cadherin and N-cadherin. \textbf{D-F}: Corresponding simulations. A-C from ref. \cite{Hayashi.n04}.}\label{fig:nocad} \end{figure}

The mutant $C$ cell in Fig.~\ref{fig:nocad}A does not express E-cadherin, and it lacks adherens junctions at the interfaces with the $P$ cells \cite{Hayashi.n04}. 
To simulate it,  it would seem natural to suppress  $J_{E}$ at all interfaces, that is, 
$J_{cP}=0$ and $J_{cC}=J_{N}$. With this assumption, we obtain the correct topology, which is the same as in the wildtype; however, the simulated geometry (not shown) is also the same as the wildtype, while the experiment is significantly different (Fig.~\ref{fig:nocad}A). If we rather assume that $C$-$C$ adhesion is unchanged by this mutation ($J_{cC}=J_{CC}$), we obtain a good agreement (Fig.~\ref{fig:nocad}D).
  
\begin{figure}
\includegraphics{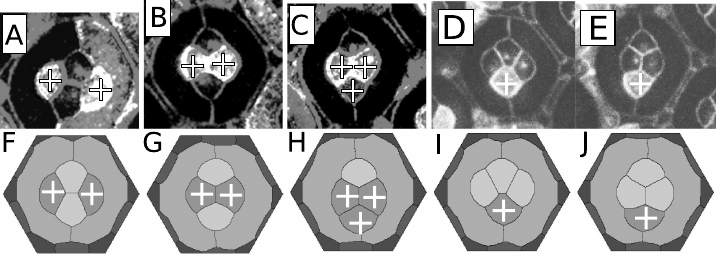}
\caption{E-cadherin overexpression.
Mutant cells are indicated with a {}``+''. \textbf{A-E}: Experimental pictures, A-C from ref. \cite{Hayashi.n04}. \textbf{F-J}: Corresponding simulations.}\label{fig:Ecad} 
\end{figure}

E-cadherin overexpression in $C$ cells (but not in $P$ cells) significantly affects the pattern, yielding a coexistence of different topologies: in Fig.~\ref{fig:Ecad}A and B,  the same cells are mutants, but the topologies differ; the same holds for Fig.~\ref{fig:Ecad} D and E. 
We predict the observed topologies (all stable) and, qualitatively, the geometries (Fig.~\ref{fig:Ecad}F-J) when
we increase the $C$-$P$ cell adhesion from $J_{CP}=J_{E}=150$ to $J_{cP}=300$;
while we find that the adhesion between a wildtype and mutant $C$ cells should not change, $J_{cC}=J_{CC}=J_{E}+J_{N}$, we should change it if both are mutants,  $J_{cc}=350+J_{N}$. Since E-cadherin overexpression in $P$ cells rarely induces geometrical or topological changes \cite{Hayashi.n04}, we do not change their adhesion values.

\subsection{E- and N-cadherin mutants} 
  
We predict the effect of both E-cadherin and N-cadherin missing in $C$ cells by setting $J_{cC}=J_{cP}=J_{cc}=0$. 
Mutant $C$ cells do not adhere to any of their neighbors, Fig.~\ref{fig:nocad}E-F:  intercellular space becomes visible between the cells, and the cells have shrunken.
This agrees well with experiments, where mutant $C$ cells lose the apical contacts with their neighbors (Fig.~\ref{fig:nocad}B-C). 

\section{Discussion}

\subsection{Constant and variable tension models}
 
When surface tension is a constant model parameter, only modified by adhesion, the surface mechanics are soap-bubble-like: minimization of the interfaces with cell type dependent weights \cite{Chichilnisky.jtb86,Graner.jtb93a,Graner.prl92,Glazier.pres93,Brodland.jbe00}. This model proves to be insufficient here. However, in studies focussing on larger aggregates ($10^{2}$ to $10^{4}$ cells) \cite{Graner.jtb93a,Brodland.jbe00,Kafer.pcb06}, constant surface tension was sufficient to explain tissue rounding and cell sorting, and even {\it Dictyostelium} morphogenesis \cite{Maree.pnas01}. This constant tension model catches two important features of tissues of adherent cells: first, cells tile the space without gaps or overlap; second, the interface between cells is under (positive) tension, which implies for instance that  three-cell vertices are stable, unlike four-cell ones, \cite{Plateau.setl73,Weaire.pf99} and thus severely constrains the possible topologies \cite{Weaire.pf99}.

In the present example of retina development, we show that interfacial tension should be variable, as described in a second model \cite{Graner.jtb93b,Ouchi.pa03}. Tension results from a adhesion-driven extension of cell-cell interfaces, balanced by an even larger cortical tension (eq.~\eqref{eq:tension}). It explains correctly the topologies of many observations, and correctly simulates the geometries. It requires more free parameters; but they are tested against many more experimental data; and their origins, signs and variations are biologically relevant \cite{Lecuit.nrmc07}. 

Adding more refinements (and thus more free parameters) would be easy, but does not seem necessary to describe the equilibrium shape of ommatidial $C$ cells. The parameters should not be taken as quantitative predictions, since {\it in vivo} biophysical measurements to compare them to are lacking. 

\subsection{Adhesion}

By adjusting a set of $6$ independent free parameters in this variable tension model, we obtain topological and geometrical agreement between the  simulations and the pictures of $16$ different situations: the wildtype (Fig.~\ref{fig:wildtype}), the six topologies observed in the \emph{Roi}-mutants (Fig.~\ref{fig:Roi} and \prettyref{sfig:Roi}); as well as the nine cadherin deletion mutants (Figs.~\ref{fig:Ncad}A-F, \ref{fig:nocad}) by setting the corresponding parameter to zero.

We also simulate $7$ cadherin overexpression mutants, by re-adjusting the corresponding parameter 
(Figs.~\ref{fig:Ncad}G,H, \ref{fig:Ecad}): adhesion is increased. The strongest increases are found when two overexpressing cells touch: this corresponds to the idea that the adhesion strength depends on the availability of cadherin molecules in both adhering cells.

We found two cases where a mutation does not seem to change the adhesion strength: first, when deleting E-cadherin from one $C$ cell, its adhesion with a normal $C$ cell is unchanged (Fig.~\ref{fig:nocad}D); second, we rarely observed shape changes in E-cadherin overexpressing $P$ cells in experiments (c.f. \cite{Hayashi.n04}). 

Indeed, while a linear relation between cadherin expression and adhesion strength has been found {\it in vitro} \cite{Foty.db05}, this need not be true {\it in vivo}, since cells have many more ways to regulate protein levels. These exceptions, thus, do not contradict the conclusion that the shapes observed in mutants are the effect of altered adhesion: an increase in the case of overexpression, a decrease in the case of deletion.

\subsection{Cortical tension} 

In the variable tension model, the perimeter modulus $\lambda_{P}$ and the target perimeter $P_{0}$ reflect the role of the cortical cytoskeleton. The target perimeter is always smaller than the perimeter, therefore the interfacial tension $\gamma_{ij}$ (eq.~\eqref{eq:tension}) is always positive, else the cell would be unstable and fall apart or disappear. The cortex of the simulated cells is contractile, and generates tension. This tension depends on the perimeter $P$ of the cell, which length depends on the cell's shape, which in turn depends on the tension: there is a feedback between tension and shape, and thus between each cell and its neighbours.

To understand the effect of this feedback, let us consider the wildtype ommatidium. We assume that the four $C$ cells have equal adhesion properties. The tension at the interfaces between the two $P$ cells pulls at the polar and equatorial $C$ cell. When the tension is constant, these cells will therefore be pulled apart (Fig.~\ref{fig:tension}A): the cells do not react on their deformation. When the tension, however, depends on the cell's perimeter, pulling at those cells deforms them, and increases their tension: energy minimization thus requires that they stay in contact.

The prediction that cytoskeletal contractility is essential for the establishment of cell shape should be tested, e.g. by treating the cells with cytoskeletal inhibitors \cite{Raucher.bj99,BarZiv.pnas99}, or genetically modifying the cytoskeleton. Since the cytoskeleton has multiple functions that could interfere with adhesion (c.f. \cite{Geisbrecht.ncb02, Gumbiner.nrmc05}), the results will be difficult to interpret. Preliminary experimental results (not shown) do indicate that genetically disturbing Rho-family GTPases influences the cell shape. 
The role of the cytoskeleton has been confirmed in various tissues and organisms (see \cite{Lecuit.nrmc07,Schock.arcd02} for reviews). We here present a computational framework able to test this hypothesis, which can be extended to other tissues, ranging from patterns of few cells to large-scale aggregates.

\section{Methods}

\subsection{Experiments}

Retinas were stained and analyzed as described in \cite{Hayashi.n04} and \cite{Hayashi.db98}. In short, cell contours were
visualized by staining either with cobalt sulfide (Fig.~\ref{fig:Roi} and \prettyref{sfig:Roi}), or with the antibodies against DE-cadherin, DN-cadherin (referred to as E- resp. N-cadherin in the rest of the text),
$\beta$-catenin or $\beta$-spectrin.
Rough eye ({\it Roi}) flies were used to examine the topology and geometry
of variable
number of cone cells. The effect of eliminating or overexpressing
cadherin molecules was studied in mosaic retinas composed of wild-type
and mutant cells (see \cite{Hayashi.n04}).
 We examined more than five retinas in each experiment. Thus at least
several hundreds of ommatidia ($> 500$) were examined for the wildtype
and each mutation, except the E- and N-cadherin overexpression, in
which case approximately $100$ ommatidia were examined each. Some pictures used for the analysis were published previously \cite{Hayashi.n04,Hayashi.db98}.

\subsection{Model simulations}

The cellular Potts model \cite{Graner.prl92,Glazier.pres93} is a standard algorithm to simulate variable cell shape, size and packing \cite{Maree.scbm07}. Its use in biology is motivated by the capability to handle irregular, fluctuating interfaces (c.f. \cite{Mombach.prl95}); the pixelisation induced by the calculation lattice can be chosen to correspond to the pixelisation in the experimental images. 

Each cell is defined as a certain set of pixels, here on a 2D square lattice; their number defines the cell area $A$. The cell shapes change when one pixel is attributed to one cell instead of another. Our field of simulation for one ommatidium is a hexagon with sides of approximately $100$ pixels (its surface is $A_{hex}=25160$ pixels, about the same as in experimental pictures). We use periodic boundary conditions, as if we were simulating an infinite retina with identical ommatidia. 
Initially, the whole hexagon is filled with cells, approximately at the right positions (\prettyref{sfig:Initial-conditions}). We treat bristle cells as tertiary pigment cells: both are situated at the edge of three ommatidia. These initial conditions, with an unstable
$n$-cell vertex in the middle, do not fix the final configuration in advance. 
Simulations can be started with different seeds of the random number generator, to explore whether multiple solutions are possible. 

Shape is relaxed in order to decrease the energy ${\cal E}$, eq. \eqref{eq:standard hamiltonian}   \cite{Graner.prl92} or eq. \eqref{eq:adhesion hamiltonian} \cite{Ouchi.pa03}.
The algorithm to minimize ${\cal E}$ uses Monte Carlo sampling and the Metropolis algorithm, as follows. We randomly draw (without replacement) a lattice pixel, and one of its eight neighboring pixels. If both pixels belong different cells, we try to copy the state of the neighboring pixel to the first one. If the copying diminishes ${\cal E}$, we accept it; and if it increases ${\cal E}$, we accept it with probability $P=\exp\left(-\Delta {\cal E}/T\right)$. Here $\Delta {\cal E}$ is the difference in ${\cal E}$ before and after the considered copying. The prefactor $T$ is a 
fluctuation (random copying) allowance:  it determines the extent of energy-increasing copy events, leading to membrane fluctuations \cite{Mombach.prl95}. Since all energy parameters are scalable with the fluctuation allowance $T$, we can fix it without loss of generality; for numerical convenience we choose numbers of order of a hundred.

We define one Monte Carlo time step (MCS) as the number of random drawings equal to the number of lattice pixels. 
It takes  approximately $600$ to $4000$ MCS to attain a shape that does not evolve anymore, that is, in mechanical equilibrium where stresses are balanced. We run the simulation much longer (up to $10^{6}$ MCS) to test if topological changes occur.

To avoid possible effects of lattice anisotropy on cell shapes, we compute $P$ and ${\cal E}$ by including interactions up to the 20 next next nearest neighbours \cite{Holm.pra91}. All perimeters indicated here are corrected by a suitable prefactor $10.6$ to ensure that a circle with an area of $A$ pixels has a perimeter $2\sqrt{\pi A}$ \cite{Raufaste.rie07}.

In experiments, interstitial fluid is present in small amount, and cells can lose contact (Fig.~\ref{fig:nocad} B,C).
To simulate it in our 2D model, at each MCS we randomly choose one pixel at a cell interface and change its state into `intercellular space' (a state without adhesion, nor area and perimeter constraints).
In addition, we choose the sum of all  cells target areas to be less than the total size of the hexagonal simulation field ($\sum_{cells}A_{0i}=0.95\; A_{hex}$, \prettyref{tab:Simulation-parameters}).
Only when cells lose adhesion ($J=0$) do we actually observe  intercellular space in simulations (Fig.~\ref{fig:nocad}E,F).

We try different parameters and adjust them to improve the visual agreement (`eyeballing') between   simulated and experimental pictures.
To estimate our uncertainty, we note that $5-10$\% changes in the values of the adhesion parameters do not yield visible changes in the geometry, while $10-30$\% changes do; see \prettyref{sfig:sens2} for an example of the determination of $J_{Cp}$.

\subsection{Images}

Once we simulate the correct topology, 
we measure the contact angles of straight lines fitted through the interfaces that meet in the vertex. The line should be long enough to avoid grid effects; we fit a straight line using the first $15$ first-order neighboring sites. Since the simulated cells show random fluctuations, statistics are obtained by measuring the contact angles several times during the simulation, or in simulations with different random number seeds.
 
\begin{figure}
\includegraphics{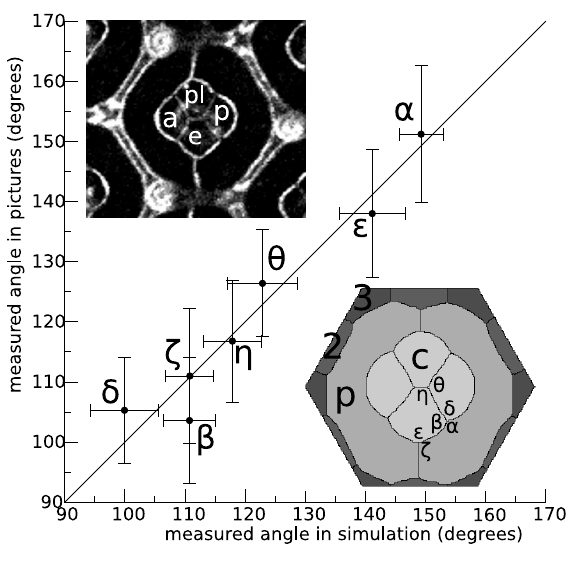}
\caption{Wildtype. 
Contact angles measured in experiments and in simulations are plotted as average $\pm$ statistical standard deviation; the straight line represents $y=x$.
Inset left: an ommatidium stained for E-cadherin; anterior (a), posterior (p), polar (pl) and equatorial (e) cone cells. Inset right: variable tension model simulation, with cone cells (C), and primary (P), secondary (2) and tertiary (3) pigment cells. One ommatidium contains four times the angles $\alpha$, $\beta$, $\delta$, $\eta$, and $\zeta$, and two times $\epsilon$ and $\theta$. }\label{fig:wildtype}
\end{figure}

In experimental pictures, we measure contact angles in $22$ wildtype ommatidia by hand, aided by the program ImageJ \cite{Rasband.web05}. Ommatidia have two axes of symmetry, and we consider the ommatidia to consist of four equal quarters, which gives us $88$ measurements for each angle (and $44$ measurements of the angles that are intersected by the axes of symmetry). 
The variation between different wildtype ommatidia is larger than in simulations (Fig.~\ref{fig:wildtype}). In mutant ommatidia, the error bar is even larger, so that we did not attempt at any quantitative comparison.

\begin{acknowledgments}
We thank Simon Cox for Surface Evolver calculations on soap bubble clusters, Christophe Raufaste for discussions on the computational methods, Sascha Hilgenfeldt for interesting discussions, and Yohanns Bella\"iche for critical reading of the manuscript.
We thank T. Uemura, H. Oda, U. Tepass, G. Thomas, B. Dickson, P.
Garrity, the Bloomington Drosophila Stock Center and the Developmental
Studies Hybridoma Bank for fly strains and/or antibodies, and K. Saigo for use of facilities. T.H. was supported by a research
fellowship from the Japan Society for the Promotion of Science for
Young Scientist.

\end{acknowledgments}

%% Add your bibliography items here. PNAS requires that bibliography items %% be entered directly into the article rather than called from a BibTeX %% environment. Contact pnas@nas.edu if you need assistance with your %% bibliography.

% Sample bibliography item in PNAS format: %% \bibitem{Ch} D. Chae (2003) {\it Nonlinearity} {\bf 16}, 479-495. %% \bibitem{in-text reference} Author Names (year published) %% {\it Journal Name} {\bf Volume #}, start page-end page 

\end{article}
%%%%%%%%%%%%%%%%%%%%%%%%%%%%%%%%%%%%%%%%%%%%%%%%%%%%%%%%%%%%%%%% 

\clearpage

\begin{figure}
\captionsetup{labelformat=support}
\caption{\label{sfig:confocal}Confocal microscopy images of the \emph{Drosophila} retina. The pictures measure $100\,\mu\mbox{m}\times100\,\mu\mbox{m}$. a) $\beta$-catenin, a component of the adherens junction, is stained green. Nearly all catenin fluorescence between the cone cells is seen in a layer of $1.26\,\mu\mbox{m}$ thick. One ommatidium consists of four cone cells (c) and two primary pigment cells (p), surrounded by six secondary (2) and three tertiary (3) pigment cells and three bristle cells (b). In this particular ommatidium, one bristle cell is replaced by a tertiary pigment cell. The cone cells can be subdivided into a polar (pl), equatorial (eq), anterior (a) and posterior (po) cone cell, according to their position. b) N-cadherin fluorescence in the same plane of focus. N-cadherin is restricted to the cone - cone interfaces.}
\end{figure}

\begin{figure}
\captionsetup{labelformat=support}
\caption{\label{sfig:Initial-conditions}Initial conditions of each simulation with four cone cells and two primary pigment cells (left), and six cone cells and three primary pigment cells (right). Periodic boundary conditions imply that the secondary pigment cells (purple) and tertiary pigment cells (red) that are marked with the same symbol, are treated as parts of the same cell.}
\end{figure}

\begin{figure}
\captionsetup{labelformat=support}
\caption{\label{sfig:Roi}Experiments and simulations showing ommatidia with 2 (A,D), 3 (B,E) and 5 (C,F) cone cells. A-C from ref. \cite{Hayashi.n04}.} \end{figure}

\begin{figure}
\captionsetup{labelformat=support}
\caption{\label{sfig:sens2}
Determination of the  adhesion between  cone cells and two N-cadherin mis-expressing pigment cells. Simulations are shown with values $J_{Cp}=150$ (A), 600 (B), 700 (C), 750 (D), 800 (E), 850 (F). 
(A) corresponds to wildtype, (D) corresponds best to the mis-expression experiment (Fig. \ref{fig:Ncad}G).}
\end{figure} 

\begin{table}
\captionsetup{labelformat=support}
\caption{\label{tab:Simulation-parameters}Simulation parameters of the wildtype in the variable tension model.
 \textparagraph: Free parameter adjusted to compare to wildtype observation.
  \textdagger: Parameters which value has little effect on the images.
 \textdaggerdbl: Parameter of no effect on the images, since cone cells  almost never touch  secondary or tertiary pigment cells.
Target perimeters are expressed as a factor times the perimeter of a circle having the specific target area. E.g. a prefactor of $1$ indicates that the target perimeter of the cell equals the perimeter if the cell is round and has an area equaling its target area. A cell with a prefactor $>1$ (like the primary pigment cells) can deviate much from a round shape.  Abbreviations: $N$, N-cadherin; $E$, E-cadherin;  $C$, cone cell;  $P$, primary pigment cell; 2, secondary pigment cell; 3, tertiary pigment cell.
} 
\end{table}

%% Adding Figure and Table References
%% Be sure to add figure and table references after \end{article} %% and before \end{document}

%% For figures, put the caption below the illustration. 

%% \begin{figure}
%% \caption{Almost Sharp Front}\label{afoto} %% \end{figure}

%% For Tables, put caption above table
%% Caption should start with a capital letter, continue with lower case %% and not have a period at the end
%% Using @{\vrule height ?? depth ?? width0pt} in the preamble will %% keep that much space between every line in the table. 

%% \begin{table}
%% \caption{Repeat length of longer allele by age of onset class} %% \begin{tabular}{@{\vrule height 10.5pt depth4pt width0pt}lrcccc} %% table text
%% \end{tabular}
%% \end{table}

%% For two column figures and tables, use the following: 

%% \begin{figure*}
%% \caption{Almost Sharp Front}\label{afoto} %% \end{figure*}

%% \begin{table*}
%% \caption{Repeat length of longer allele by age of onset class} %% \begin{tabular}{ccc}
%% table text
%% \end{tabular}
%% \end{table*}


\begin{thebibliography}{10}

\bibitem{Steinberg.s63}
Steinberg, M.~S. (1963) {\em Science\/} {\bf 141}, 401--408.

\bibitem{Niewiadomska.jcb99}
Niewiadomska, P., Godt, D., \& Tepass, U. (1999) {\em J. Cell Biol.\/} {\bf
  144}, 533--547.

\bibitem{Foty.ijdb04}
Foty, R.~A. \& Steinberg, M.~S. (2004) {\em Int. J. Dev. Biol.\/} {\bf 48},
  397--409.

\bibitem{Lecuit.tcb05}
Lecuit, T. (2005) {\em Trends Cell Biol.\/} {\bf 15}, 34--42.

\bibitem{Classen.dc05}
Classen, A.~K., Anderson, K.~I., Marois, E., \& Eaton, S. (2005) {\em Dev.
  Cell\/} {\bf 9}, 805--817.

\bibitem{Gumbiner.nrmc05}
Gumbiner, B.~M. (2005) {\em Nat. Rev. Mol. Cell Biol.\/} {\bf 6}, 622--634.

\bibitem{Carthew.cogd05}
Carthew, R.~W. (2005) {\em Curr. Opin. Genet. Dev.\/} {\bf 15}, 358--363.

\bibitem{Wolff.ddm93}
Wolff, T. \& Ready, D. (1993) In {\em The Development of {\emph{Drosophila
  melanogaster}}\/}, eds. Bate, M. \& Martinez~Arias, A. (Cold Spring Harbor
  Laboratory Press), pp. 1277--1325.

\bibitem{Hayashi.n04}
Hayashi, T. \& Carthew, R.~W. (2004) {\em Nature\/} {\bf 431}, 647--652.

\bibitem{Plateau.setl73}
Plateau, J. (1873) {\em Statique exp\'erimentale et th\'eorique des liquides
  soumis aux seules forces mol\'eculaires\/}, volume~1 (Gauthier-Villars,
  Paris).

\bibitem{Weaire.pf99}
Weaire, D. \& Hutzler, S. (1999) {\em The Physics of Foams\/} (Oxford
  University Press).

\bibitem{Thompson.ogaf42}
Thompson, D. (1942) {\em On Growth and Form: A New Edition\/} (Cambridge
  University Press).
\newblock Reprinted by Dover Publications, New York, 1992.

\bibitem{Chichilnisky.jtb86}
Chichilnisky, E.~J. (1986) {\em J. theor. Biol.\/} {\bf 123}, 81--101.

\bibitem{Lecuit.nrmc07}
Lecuit, T. \& Lenne, P.~F. (2007) {\em Nat. Rev. Mol. Cell Biol.\/} {\bf 8},
  633--644.

\bibitem{Graner.prl92}
Graner, F. \& Glazier, J.~A. (1992) {\em Phys. Rev. Lett.\/} {\bf 69},
  2013--2016.

\bibitem{Glazier.pres93}
Glazier, J.~A. \& Graner, F. (1993) {\em Phys. Rev. E\/} {\bf 47}, 2128--2154.

\bibitem{Graner.jtb93a}
Graner, F. (1993) {\em J. theor. Biol.\/} {\bf 164}, 455--476.

\bibitem{Sheetz.tcb96}
Sheetz, M.~P. \& Dai, J. (1996) {\em Trends Cell Biol.\/} {\bf 6}, 85--89.

\bibitem{Raucher.bj99}
Raucher, D. \& Sheetz, M.~P. (1999) {\em Biophys. J.\/} {\bf 77}, 1992--2002.

\bibitem{Dai.bj99}
Dai, J. \& Sheetz, M.~P. (1999) {\em Biophys. J.\/} {\bf 77}, 3363--3370.

\bibitem{Morris.jmb01}
Morris, C.~E. \& Homann, U. (2001) {\em J. Membr. Biol.\/} {\bf 179}, 79--102.

\bibitem{Thoumine.ebj99}
Thoumine, O., Cardoso, O., \& Meister, J.~J. (1999) {\em Eur. Biophys. J.\/}
  {\bf 28}, 222--234.

\bibitem{Graner.jtb93b}
Graner, F. \& Sawada, Y. (1993) {\em J. theor. Biol.\/} {\bf 164}, 477--506.

\bibitem{Ouchi.pa03}
Ouchi, N.~B., Glazier, J.~A., Rieu, J.~P., Upadhyaya, A., \& Sawada, Y. (2003)
  {\em Physica A\/} {\bf 329}, 451--458.

\bibitem{Maree.scbm07}
Mar{\'e}e, A. F.~M., Grieneisen, V.~A., \& Hogeweg, P. (2007) In {\em Single
  Cell Based Models in Biology and Medicine\/}, eds. Anderson, A. R.~A.,
  Chaplain, M. A.~J., \& Rejniak, K.~A. (Birkh{\"a}user-Verlag, Basel), pp.
  107--136.

\bibitem{Langmuir.jcp33}
Langmuir, I. (1933) {\em J. Chem. Phys.\/} {\bf 1}, 756--776.

\bibitem{Brodland.jbe00}
Brodland, G.~W. \& Chen, H.~H. (2000) {\em J. Biomech. Eng.\/} {\bf 122},
  402--407.

\bibitem{Kafer.pcb06}
K\"{a}fer, J., Hogeweg, P., \& Mar\'{e}e, A.~F. (2006) {\em PLoS. Comput.
  Biol.\/} {\bf 2}, e56.

\bibitem{Maree.pnas01}
Mar\'{e}e, A. F.~M. \& Hogeweg, P. (2001) {\em Proc. Natl. Acad. Sci. U.S.A.\/}
  {\bf 98}, 3879--3883.

\bibitem{Foty.db05}
Foty, R.~A. \& Steinberg, M.~S. (2005) {\em Dev. Biol.\/} {\bf 278}, 255--263.

\bibitem{BarZiv.pnas99}
Bar-Ziv, R., Tlusty, T., Moses, E., Safran, S.~A., \& Bershadsky, A. (1999)
  {\em Proc. Natl. Acad. Sci. U.S.A.\/} {\bf 96}, 10140--10145.

\bibitem{Geisbrecht.ncb02}
Geisbrecht, E.~R. \& Montell, D.~J. (2002) {\em Nat. Cell Biol.\/} {\bf 4},
  616--620.

\bibitem{Schock.arcd02}
Schock, F. \& Perrimon, N. (2002) {\em Annu. Rev. Cell Dev. Biol.\/} {\bf 18}, 
463--493

\bibitem{Hayashi.db98}
Hayashi, T., Kojima, T., \& Saigo, K. (1998) {\em Dev. Biol.\/} {\bf 200},
  131--145.

\bibitem{Mombach.prl95}
Mombach, J.~C., Glazier, J.~A., Raphael, R.~C., \& Zajac, M. (1995) {\em Phys.
  Rev. Lett.\/} {\bf 75}, 2244--2247.

\bibitem{Holm.pra91}
Holm, E.~A., Glazier, J.~A., Srolovitz, D.~J., \& Grest, G.~S. (1991) {\em
  Phys. Rev. A\/} {\bf 43}, 2662--2668.

\bibitem{Raufaste.rie07}
Raufaste, C. (2007) {\em Rh\'{e}ologie et imagerie des \'{e}coulements 2D
  de mousse\/}.
\newblock Ph.D. thesis, Universit\'{e} Joseph Fourier, Grenoble I.

\bibitem{Rasband.web05}
Rasband, W. (2005) Image{J} 1.34s.
\newblock http://rsb.info.nih.gov.ij/.

\end{thebibliography}
\end{document}